\documentclass[a4paper,11pt]{article}
\usepackage{jcappub} 
\makeatletter
\gdef\@fpheader{}
\makeatother

\title{\boldmath Birth of baby universes from gravitational collapse in a modified-gravity scenario}

\newcommand{\R}{\mathcal{R}}

\author[a]{Andreu Mas\'o-Ferrando,}
\affiliation[a]{Departament de F\'{i}sica Te\`{o}rica and Instituto de F\'isica Corpuscular (IFIC), Centro Mixto Universitat de Val\`encia - Consejo Superior de Investigaciones Cient\'ificas (CSIC).
    Universitat de Val\`encia, Burjassot-46100, Val\`encia, Spain}
\author[b,c]{Nicolas Sanchis-Gual,}
\affiliation[b]{Departament d'Astronom\'{\i}a i Astrof\'{\i}sica, Universitat de Val\`encia,
    Dr. Moliner 50, 46100, Burjassot (Val\`encia), Spain}	
\affiliation[c]{Departamento  de  Matem\'{a}tica  da  Universidade  de  Aveiro  and  Centre  for  Research  and  Development in  Mathematics  and  Applications  (CIDMA),  Campus  de  Santiago,  3810-183  Aveiro,  Portugal} 

\author[b,d]{Jos\'e A. Font,}
\affiliation[d]{Observatori Astron\`omic, Universitat de Val\`encia, Catedr\`atic 
    Jos\'e Beltr\'an 2, 46980, Paterna (Val\`encia), Spain}
\author[a,e]{and Gonzalo J. Olmo}
\affiliation[e]{Universidade Federal do Cear\'a (UFC), Departamento de F\'isica,\\ Campus do Pici, Fortaleza - CE, C.P. 6030, 60455-760 - Brazil.}

\emailAdd{andreu.maso@uv.es}
\emailAdd{nicolas.sanchis@uv.es}
\emailAdd{j.antonio.font@uv.es}
\emailAdd{gonzalo.olmo@uv.es}

\abstract{We consider equilibrium models of spherical boson stars in Palatini $f(\R)=\R+\xi \R^2$ gravity and study their collapse when perturbed. The Einstein-Klein-Gordon system is solved using a recently established correspondence in an Einstein frame representation. We find that, in that frame, the endpoint is a nonrotating black hole surrounded by a quasi-stationary cloud of scalar field. However, the dynamics in the $f(\R)$ frame is dramatically different. The innermost region of the collapsing object exhibits the formation of a finite-size, exponentially-expanding {\it baby universe} connected with the outer (parent) universe via a minimal area surface (a throat or umbilical cord). Our simulations indicate that this surface is  at all times hidden inside a horizon, causally disconnecting the baby universe from  observers above the horizon. The implications of our findings in other areas of gravitational physics are also discussed.}

\begin{document}
\maketitle
\flushbottom

\section{Introduction}
The formation of singularities under reasonable initial conditions in General Relativity (GR) \cite{Penrose:1969pc,Hawking:1970zqf,Senovilla:2014gza}
 has been the driving force of multiple efforts to understand the nature and implications of these pathologies and also of possible mechanisms that could avoid them. Quantum approaches  and phenomenological descriptions \cite{Brandenberger:1993ef,Khoury:2001wf,Gasperini:2002bn,Novello:2008ra,Ashtekar:2011ni} suggest that our expanding universe could come from a previously contracting phase and that geodesic completeness in black hole geometries could be restored, among other possibilities \cite{Joshi:2008zz}, via a bounce in the radial sector, leading generically to the existence of minimal nonzero bounds to the area/volume in which matter fields can be concentrated. In this sense, the classical collapse model of Oppenheimer and Snyder \cite{Oppenheimer:1939ue} offers a glimpse on how a nonsingular collapse process could proceed. The innermost region of the collapsing object could be modeled as a contracting cosmology which would bounce at a certain critical density, preventing total collapse. The evolution of the bouncing material should depend crucially on the formation or not of a horizon, because the causal structures in both cases are radically different. Without a horizon, the collapsing material should be ejected back to where it came from, though the energy scales expected in such a quantum gravity process have never been observed. If a horizon forms, the bounce should proceed much more quietly for an external observer, as the interior would be causally disconnected from it. What may happen inside is still a matter of speculation. 

In this work we explore this idea by considering the collapse of boson stars~\cite{Liebling:2012fv,Schunck:2003kk}, self-gravitating compact objects that can be constructed by minimally coupling a complex, massive scalar field to gravity and which under certain conditions imitate the phenomenology of black holes
\cite{CalderonBustillo:2020fyi,CalderonBustillo:2022cja}. Our model considers a modified gravity scenario defined by a quadratic $f(\R)$ extension of GR that is known to provide bouncing cosmological solutions \cite{Barragan:2009sq}. This theory is also intimately related to effective descriptions of nonsingular models of quantum gravity \cite{Olmo:2008nf,Delhom:2023xxp}. We find that the collapse generates a horizon but also develops a nonzero minimal area region near the center followed by an inflating bubble that represents the birth of a baby universe. The bubble expands at superluminal speed and the minimal surface is sustained by the energy density of the scalar field, which leaks in from a quasistationary cloud that remains bounded around the black hole. This exterior solution is consistent with previous results in the literature of GR \cite{Barranco:2012qs,Sanchis-Gual:2014ewa} and the numerical evolution suggests that the minimal area will decay to zero when the scalar cloud is completely absorbed by the black hole. During the stationary phase, the space-time is qualitatively in agreement with results from the loop quantization of black holes \cite{Gambini:2020nsf}, heuristic {\it black bounce} models \cite{Simpson:2018tsi,Lobo:2020ffi}, and other static solutions \cite{Olmo:2012nx,Olmo:2013gqa}, though always preserving an Euclidean topology. 

\section{Framework}
Our model is based on a metric-affine (Palatini) formulation of gravity \cite{Olmo:2011uz,Hehl:1994ue}, 
in which both the metric and the affine connection are regarded {\it a priori} as independent geometric fields. Except for the Lovelock family of theories \cite{Exirifard:2007da}, which includes GR, the metric-affine formulation yields field equations that differ from the more usual metric approach. In the case of $f(\R)$ and other theories whose gravity Lagrangian is a functional of the symmetric part of the Ricci tensor \cite{Afonso:2017bxr}, the resulting equations are of second order, free of ghost instabilities \cite{BeltranJimenez:2020sqf}, and gravitational waves in vacuum propagate at the speed of light. In addition, when minimally coupled to specific matter fields, their equations can be rewritten in the Einstein frame of an auxiliary  metric $q_{\mu\nu}$ (not necessarily conformal with $g_{\mu\nu}$) minimally coupled to a nonlinear version of the original matter source \cite{Orazi:2020mhb,Afonso:2018hyj}. This allows to attack the modified gravity problem by solving first the equations of  GR minimally coupled to a modified matter source and then transforming back to the original frame variables. The implementation of numerical methods is thus greatly simplified. For example, the action of a BS in Palatini $f(\R)$ gravity is
	\begin{equation}\label{eq:ac1}
			S_{f(\R)}=\int d^4r \sqrt{-g} \frac{f(\R)}{2 \kappa} -\frac{1}{2}\int d^4r \sqrt{-g} \left(X-2V(\Phi)\right) \, ,
	\end{equation}
where the matter sector is represented by a complex scalar field $\Phi$, with $X\equiv g^{\alpha \beta}\partial_{\alpha}\Phi^*\partial_{\beta}\Phi$, $V(\Phi)=-\mu^2 \Phi^* \Phi/2$, $\mu$ is the scalar field mass, and $\kappa=8\pi$ (in $G=c=1$ units). Here, we  define $\mathcal{R}=g^{\mu\nu}R_{\mu\nu}(\Gamma)$, with $R_{\mu\nu}(\Gamma)$ representing the Ricci tensor of a connection $\Gamma^\lambda_{\alpha\beta}$ {\it a priori} independent of the metric $g_{\mu\nu}$. Taking for concreteness $f(\R)=\R+\xi \R^2$, it can be shown \cite{Afonso:2018hyj} that the associated Einstein frame theory is 
	\begin{equation}\label{eq:ac2}
			S_{\rm EF}=\int d^4x \sqrt{-q} \frac{R}{2 \kappa}-\frac{1}{2}\int d^4x \sqrt{-q} \left(\frac{Z-\xi \kappa Z^2-2V(\Phi)}{1-8 \xi \kappa V(\Phi)} \right) \quad ,
	\end{equation}
where the kinetic term $Z\equiv q^{\alpha \beta}\partial_{\alpha}\Phi^*\partial_{\beta}\Phi$ is now contracted with the (inverse) metric $q^{\alpha \beta}$, and $R$ is the Ricci scalar of the metric  $q_{\alpha \beta}$, i.e., $R= q^{\alpha \beta}R_{\alpha \beta}(q)$. The field equations also show that $\Gamma^\lambda_{\alpha\beta}$ is the Levi-Civita connection of $q_{\mu\nu}$. We will refer to the representation \eqref{eq:ac1} as the $f(\mathcal{R})$ frame, while \eqref{eq:ac2} will be the Einstein frame. By solving the equations of \eqref{eq:ac2}, algebraic relations allow to obtain the solutions of \eqref{eq:ac1}. That will be our strategy to solve the numerical problem. In particular, for $f(\R)$ theories one finds that 
	\begin{equation}\label{eq:conformal}
		q_{\mu\nu}\equiv f_{\mathcal{R}} g_{\mu\nu} \ ,
	\end{equation}
where $f_\mathcal{R}\equiv \partial f/\partial \mathcal{R}$ can be written in terms of the matter source by virtue of the field equations,which yield the algebraic relation 
\begin{equation}\label{eq:TraceEq}
    \mathcal{R}f_{\mathcal{R}}-2f=\kappa T \ .
\end{equation}
This equation implies that $\mathcal{R}=\mathcal{R}(T)$ is, in general, a model-dependent nonlinear function of the trace $T$ of the matter stress-energy tensor. When $f(\mathcal{R})=\mathcal{R}$, one recovers the expected linear relation $\mathcal{R}=-\kappa T$.  For the quadratic model to be considered here, $f(\R)=\R+\xi \R^2$, one also finds $\mathcal{R}=-\kappa T$, but in general a nonlinear relation is expected.

For completeness, we would like to comment a bit further on the implications of Eq.(\ref{eq:conformal}), as it is important to understand the suitability of considering Palatini $f(\mathcal{R})$ gravity to study the time evolution of stellar models. Using the fact that the conformal factor $f_{\mathcal{R}}$ is a function of the matter, via the relation $\mathcal{R}=\mathcal{R}(T)$, one finds that the derivatives of $g_{\mu\nu}$ depend on derivatives of $q_{\mu\nu}$ but also on derivatives of the trace of $T$ (weighted by derivatives of $f_{\mathcal{R}}$). Consequently, this peculiarity can lead to interesting phenomenology in high-energy scenarios involving strong matter gradients, though it may also lead to undesired effects when one considers simplified stellar models in which a self-gravitating polytropic fluid is matched to an exterior Schwarzschild solution. Key in this issue is the observation that for an equation of state of the form $\rho(P)\propto P^{1/\gamma}$, where gamma is the polytropic index, radial derivatives of the energy density $\rho$ can generate divergences near the surface, defined as the region where the pressure $P\to 0$. A detailed analysis (see \cite{Barausse:2007pn,Barausse:2007ys,Barausse:2008nm} but also \cite{Olmo:2008pv}) showed  that curvature scalars diverge in that limit for  polytropic fluids with $3/2<\gamma<2$, a range that includes the relevant case $\gamma=5/3$ that describes a gas of non-relativistic degenerate fermions. This disturbing effect was used to conclude that Palatini $f(\mathcal{R})$ theories were intrinsically pathological. Yet, the analysis of \cite{Barausse:2007pn,Barausse:2007ys,Barausse:2008nm} did not use consistent junction conditions at the boundary layer that separates the interior and exterior configurations, and a more rigorous analysis based on tensorial distributions \cite{Olmo:2020fri} shows that divergences only arise if $\gamma>2$, shifting the problematic range of $\gamma$ beyond the domain of direct physical interest. On the other hand, it has been verified that boson star models in Palatini $f(\mathcal{R})$ are free from any pathologies on their outermost regions \cite{Maso-Ferrando:2021ngp} and it is expected that any self-gravitating fundamental field (either boson or fermion) will be free of the pathologies observed in polytropic models. Under this light, we conclude that the peculiarity of the field equations of Palatini theories may require in some situations the use of sources with differentiability profiles smoother than in GR, but that lack of convenience is not a solid argument to rule out such theories.  

As mentioned above, the fact that a large family of metric-affine theories, which include Palatini $f(\mathcal{R})$, can be mapped into an Einstein frame representation without introducing any new dynamical degrees of freedom puts forward that the modified dynamics of these theories is encoded in nonlinearities of the matter sector (see \cite{Orazi:2020mhb,Afonso:2018hyj,Magalhaes:2022esc}  for details and examples). This explains why the scalar degree of freedom of the scalar-tensor representation of Palatini $f(\mathcal{R})$ theories is non-dynamical. The scalar object $\phi\equiv f_\mathcal{R}$ is a model-dependent algebraic function of the matter fields governed by Eq.(\ref{eq:TraceEq}) and, therefore, it is not an arbitrary but a concrete function of $T$ once the $f(\mathcal{R})$ model is specified. That algebraic relation between frames implies that the wellposedness of the modified gravity equations is guaranteed by the wellposedness of the GR equations (Einstein frame), as long as the matrix that relates $q_{\mu\nu}$ and $g_{\mu\nu}$ (conformal in the $f(\mathcal{R})$ case) is nondegenerate and sufficiently smooth. This is guaranteed, in particular, for fundamental fields like the complex scalar considered here, which further justifies our numerical strategy to solve the associated GR problem. 

The encoding of the modified dynamics in nonlinearities of the matter sector also has deep implications when these theories are considered from an effective field theory perspective, an aspect that was analyzed in detail in \cite{BeltranJimenez:2021oaq}. For a general matter sector, the effective field theory approach implies that the predictions of Ricci-based gravity theories are degenerate with those of GR, both at the classical and quantum levels. This is a natural consequence of the peculiarities of the Palatini dynamics, which is generated by nonlinearities in the matter sector. In the effective field theory framework, that amounts to a redefinition of the effective coupling parameters because the nonlinear transformation simply modifies the coefficients of terms that were already present in the matter action. Thus, no quantum inconsistencies can be argued to invalidate these theories either \cite{Iglesias:2007nv}. If specific matter sectors are considered, as is our case, the predictions of a given Palatini theory will be different from those of GR, though the Einstein-frame representation is still very useful to solve the field equations, as emphasized above.

\section{Initial data and methodology}
To study the dynamics of the collapse of a BS we use the Baumgarte-Shapiro-Shibata-Nakamura (BSSN) formalism of Einstein's equations \cite{Baumgarte:1998te, Shibata:1995we} in the Einstein frame. Details of our specific numerical implementation, including a discussion on the convergence of our code, are provided in the appendix \ref{appendix}. The initial data describing spherically-symmetric BS in Palatini gravity were obtained in~\cite{Maso-Ferrando:2021ngp}. We focus on an unstable initial configuration that undergoes gravitational collapse, choosing a central scalar field value of $\Phi_0=0.1$ and a coupling parameter $\xi=0.1$. Results for stable configurations will be reported elsewhere. To trigger the collapse we add a 3\% perturbation to the initial radial distribution of the scalar field. This leads to a slight violation of the constraints. Even though it is larger than the discretization error, it is small enough not to substantially alter the original solution. Since our initial configurations~\cite{Maso-Ferrando:2021ngp} are obtained in polar-areal coordinates but the time evolution is done in isotropic coordinates, we perform a coordinate transformation following~\cite{Lai:2004fw}.  The polar-areal grid  is equidistant with spatial resolution $\Delta x_{\text{pa}}=0.0025$. After the transformation to isotropic coordinates a cubic-spline  interpolation is applied to have the initial configuration on a two-patch grid, with a geometrical progression in the interior part up to a given radius and a hyperbolic cosine outside (see~\cite{Sanchis-Gual:2015sxa} for details). To properly capture the highly non-linear, strong-field dynamics of the system close to the center of the star (see below) a fairly small minimum resolution is required for the logarithmic grid, namely $\Delta x_{\text{iso}}=1.25\times 10^{-3}$. With this choice, the inner boundary is placed at $x_{\text{min}}^{\text{iso}}=6.25\times 10^{-4}$ and the outer boundary at $x_{\text{max}}^{\text{iso}}=1500$, using a grid with $2\times 10^4$ zones. A Courant-condition-satisfying time step of  $\Delta t =0.3\Delta x_{\text{iso}}$ is chosen to obtain long-term stable simulations. Those are performed using an updated version of the code reported in~\cite{Escorihuela-Tomas:2017uac}. 

 \begin{figure}[t]
 \centering
\includegraphics[width=0.8\linewidth]{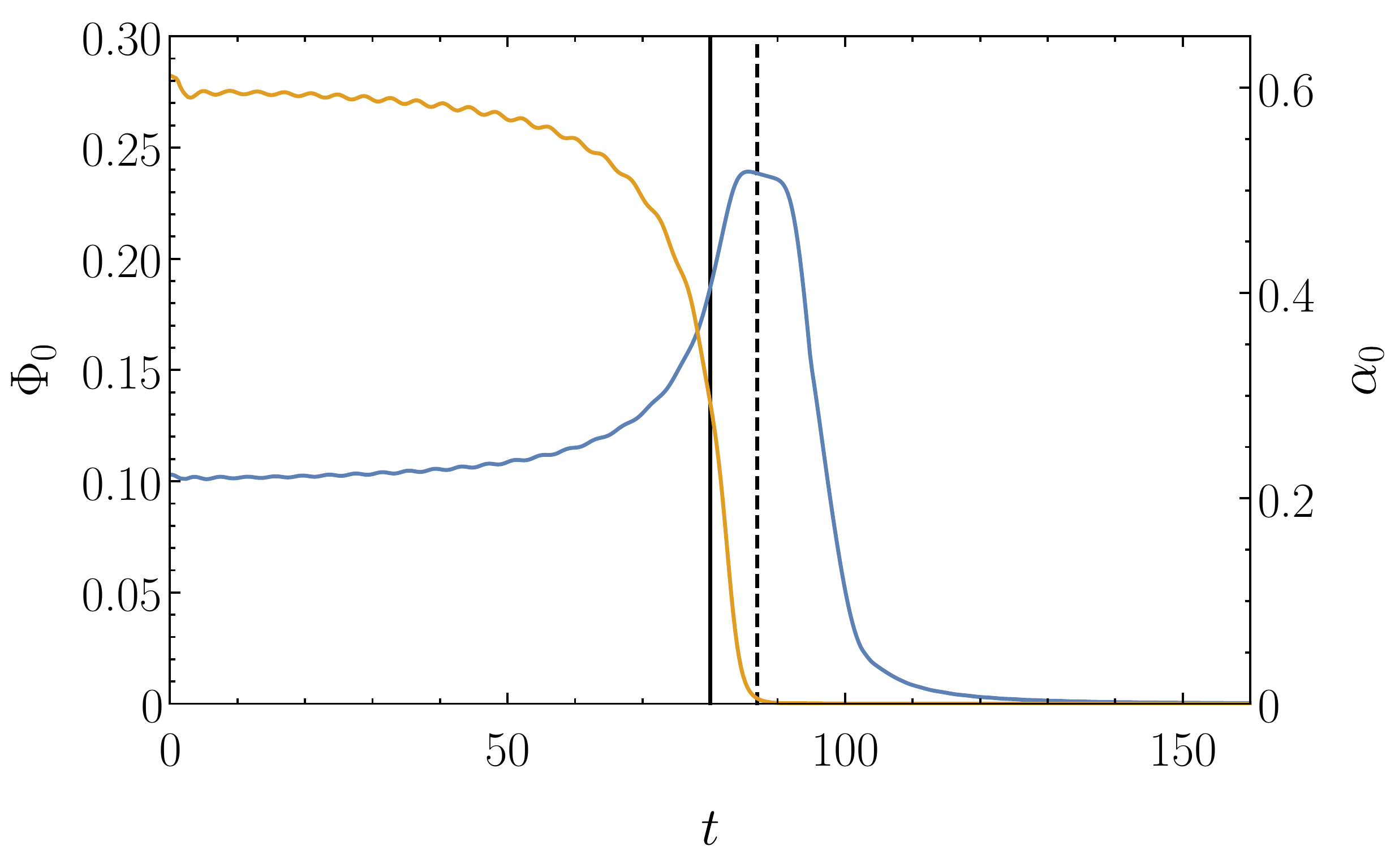}
		\caption{Time evolution of the central values of the scalar field (blue line) and of the lapse function (orange line). The black solid line indicates the instant at which the EH forms ($t\approx 80$) while the black dashed line is the time at which the AH is found ($t\approx 87$).} 
		\label{fig:plot1}
	\end{figure}

 \section{Results}
We start analyzing the dynamics of the collapse in the Einstein frame (i.e.~the GR problem \eqref{eq:ac2}).  Figure~\ref{fig:plot1} shows the evolution of the central value of the scalar field $\Phi_0$. This quantity grows up to a maximum to then decay when an apparent horizon (AH) appears. The AH is computed using the AH finder described in~\cite{Thornburg:2006zb}. The figure also depicts the instant at which the event horizon (EH) forms. The AH, defined as the outermost closed surface on which all outgoing photons normal to it have zero expansion, is a local notion and can be monitored on each time step. On the contrary, the EH is computed {\it a posteriori} tracing backwards the last trapped null geodesic \cite{Diener:2003jc}.  The AH is first found at time $t\approx 87$ and its mass, in units of $M_{\rm Pl}/\mu$, is $M_{\text{AH}}=0.61831$, slightly lower than the Misner-Sharp mass of the initial BS, $M_{\text{MS}}=0.61918$.   Figure~\ref{fig:plot1} also displays the time evolution of the central value of the lapse function, $\alpha_0$, showing the distinctive collapse-of-the-lapse once the horizon forms. The small-amplitude oscillations of $\Phi_0$ and $\alpha_0$ during the collapse are induced by the non-linearities of the $f(\mathcal{R})$ matter Lagrangian. In addition, the shift vector at the origin $\beta_0$ (not shown) attains non-zero values. The behavior of both $\alpha_0$ and $\beta_0$ reflect the singularity-avoiding slicing employed in the simulation and the presence of a singularity at the origin. 
Moreover, the metric function $q_{xx}$ grows rapidly near the center when the collapse starts, reaching values that are several orders of magnitude higher than the initial one. On the other hand, $q_{tt}$, that initially is everywhere positive, decreases 
changing sign and approaching zero from below at the center. Therefore, all metric functions mark the presence of a black hole. In the matter sector, almost all of the scalar field is swallowed by the black hole by the end of the simulation. However, a remnant of scalar field is left outside the AH in the form of a quasi-stationary long-lived cloud~\cite{Escorihuela-Tomas:2017uac,Sanchis-Gual:2017bhw}. This explains the mass disparity between $M_{\text{AH}}$ and $M_{\text{MS}}$. We note that this evolution is qualitatively identical to that of a collapsing BS in GR (without the $\xi \R^2$ term in the $f(\R)$ functional). The outcome is also a black hole whose parameters are determined by the progenitor BS model.

\begin{figure}[h]
	\centering	
    \includegraphics[width=0.8\linewidth]{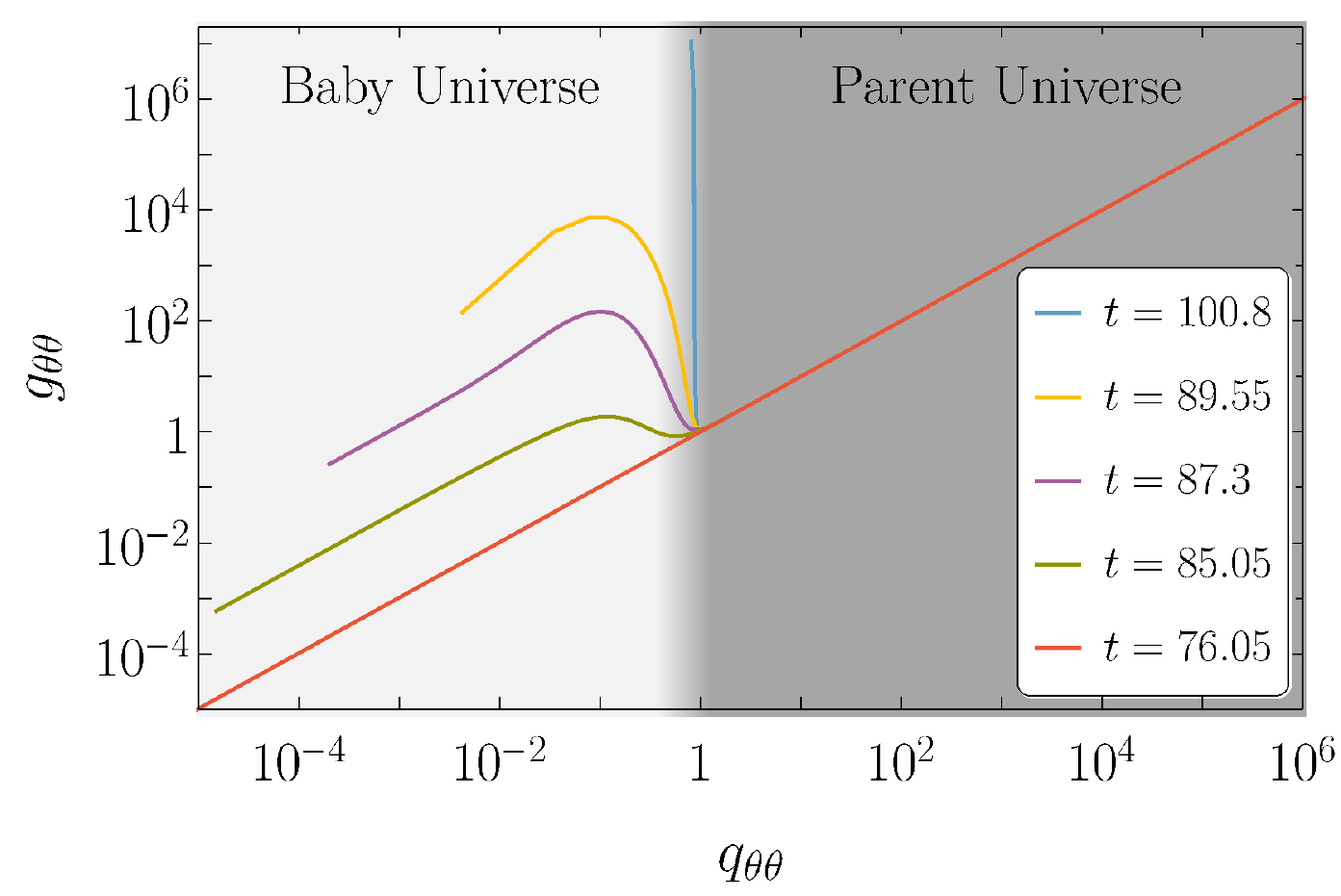}
		\caption{Relationship between the area of the two-spheres in both frames at five selected times. The background indicates the regions referred to as \textit{baby universe} and \textit{parent universe}.}
		\label{fig:plot2}
\end{figure}
 
To analyze the evolution in the $f(\R)$ frame, we need to pay special attention to the conformal factor $f_{\R}$ that relates the metrics in both frames via Eq.~\eqref{eq:conformal}. As shown in figure~\ref{fig:plot2}, at the onset and until $t\approx 80$, the area of the two-spheres of the $f(\R)$ frame, $A\sim g_{\theta \theta}$, decreases monotonically as the center\footnote{In the Einstein frame, the center is where $q_{\theta \theta}=0$.}  of the BS is approached (red curve). As the collapse proceeds and the energy density grows at the center, $f_\R$ evolves towards zero at a certain distance close to the center of the BS. As a result, a local minimum arises in $g_{\theta \theta}$ which is soon followed by a local maximum, whose height grows exponentially fast in time. The presence of a minimal two-sphere in $g_{\theta \theta}$ can be  interpreted as a cosmic bounce, i.e.~as the hypersurface that connects the contracting two-spheres (from the AH inwards) with the expanding two-spheres of the newborn universe. This baby universe is thus growing out of the patch comprised between the minimal two-sphere and the BS center.  
We will refer to the outer universe as \textit{parent universe} (PU) while the term \textit{baby universe} (BU) will be used for the inner expanding patch. 
Their corresponding areas are displayed in figure~\ref{fig:plot2}. Following~\cite{Visser:1995cc} the late-time phase of the collapse can thus be interpreted as generating a quasi-permanent inter-universe wormhole, with the bounce representing a kind of umbilical cord connecting the PU and the BU. 

One can verify that radial null geodesics between the minimal and maximal spheres follow divergent trajectories, which refocus as they go from the maximal sphere towards the center. Due to numerical limitations associated with the singularity-avoiding slicing conditions used in the Einstein frame, we can not confirm if they converge at the center. In particular, the region between the center and the maximal sphere becomes unreachable beyond  $t=91.8$.
In the time interval $t\in [84.6,91.8]$ the expansion of the BU is exponential and superluminal, always preserving the original $\mathbb{R}^4$ topology.

        \begin{figure*}[b]
			\centering
			\includegraphics[width=0.32\textwidth]{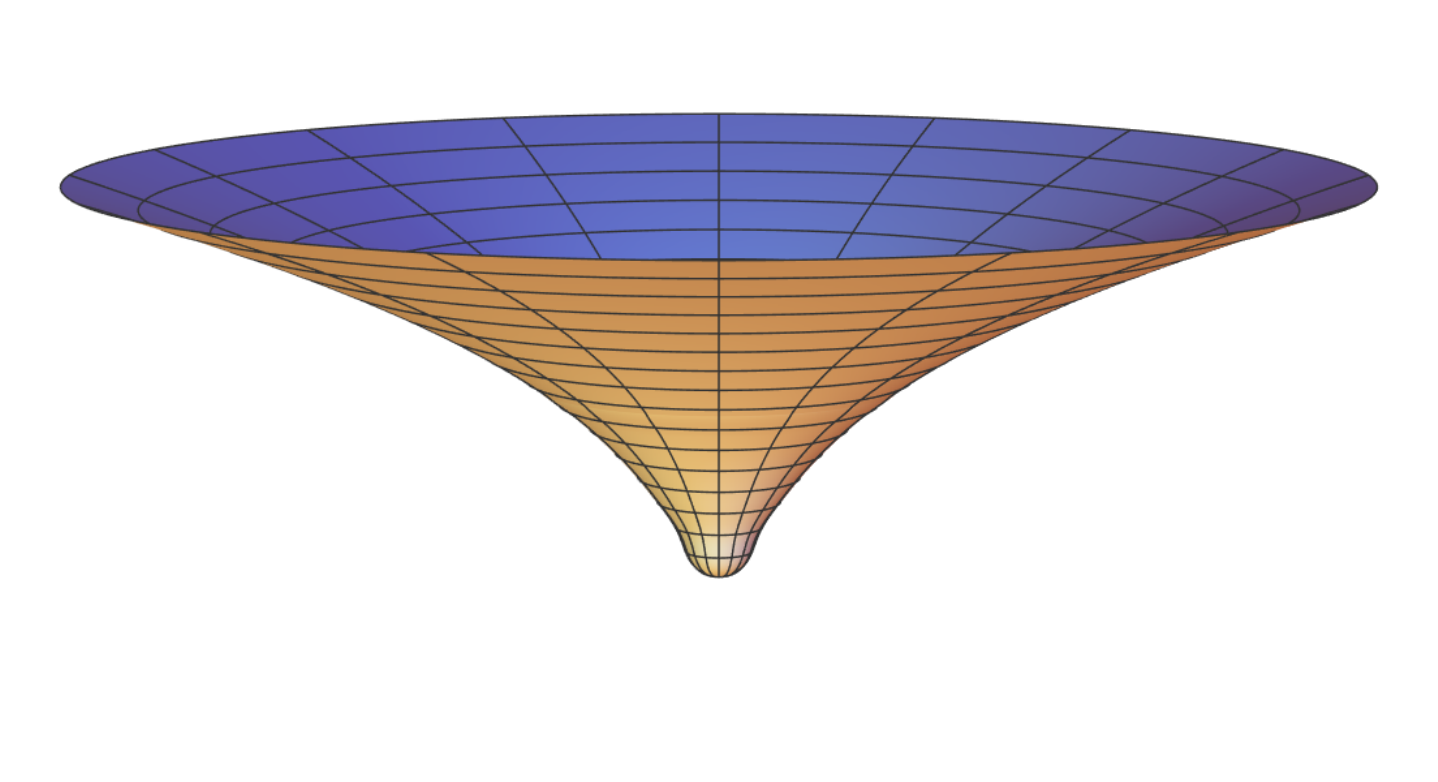}\includegraphics[width=0.32\textwidth]{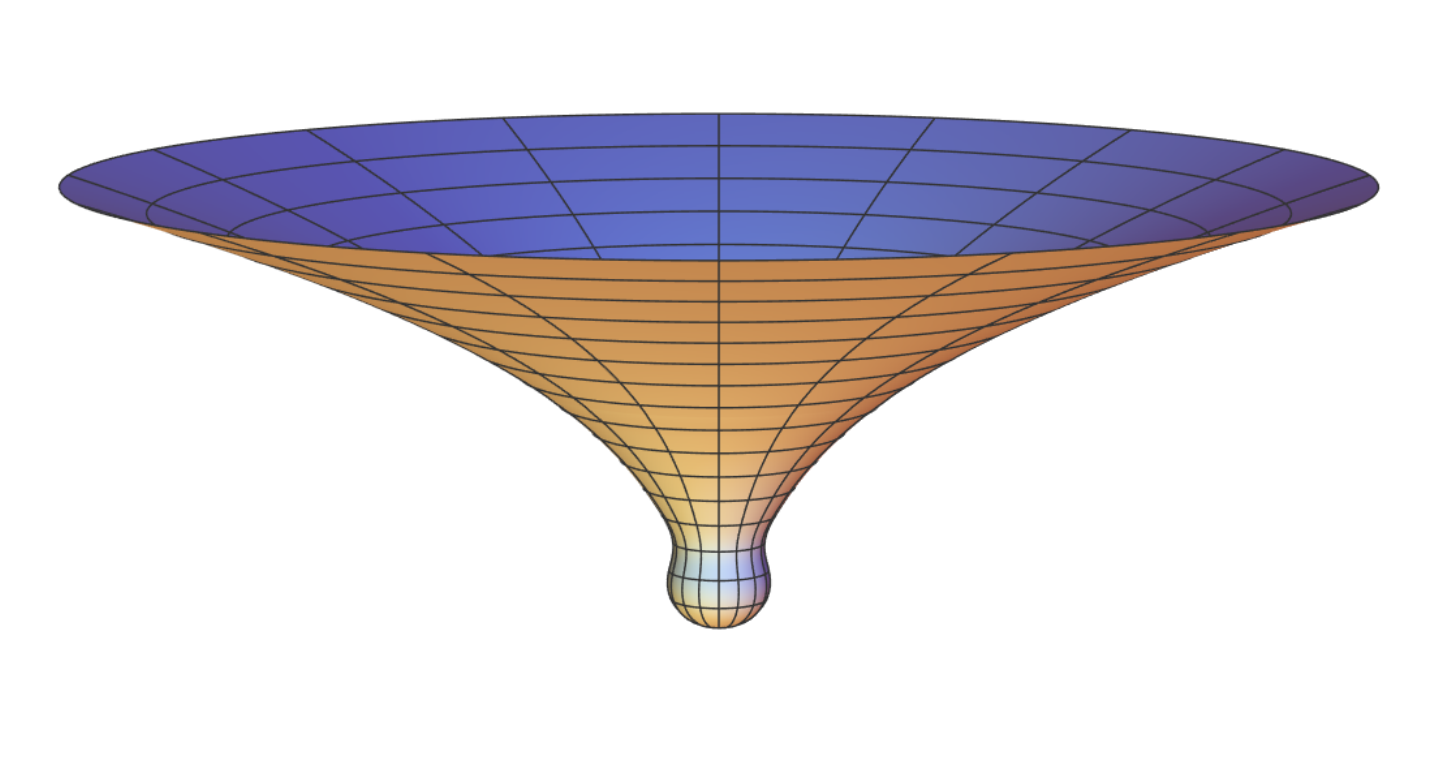}		\includegraphics[width=0.32\textwidth]{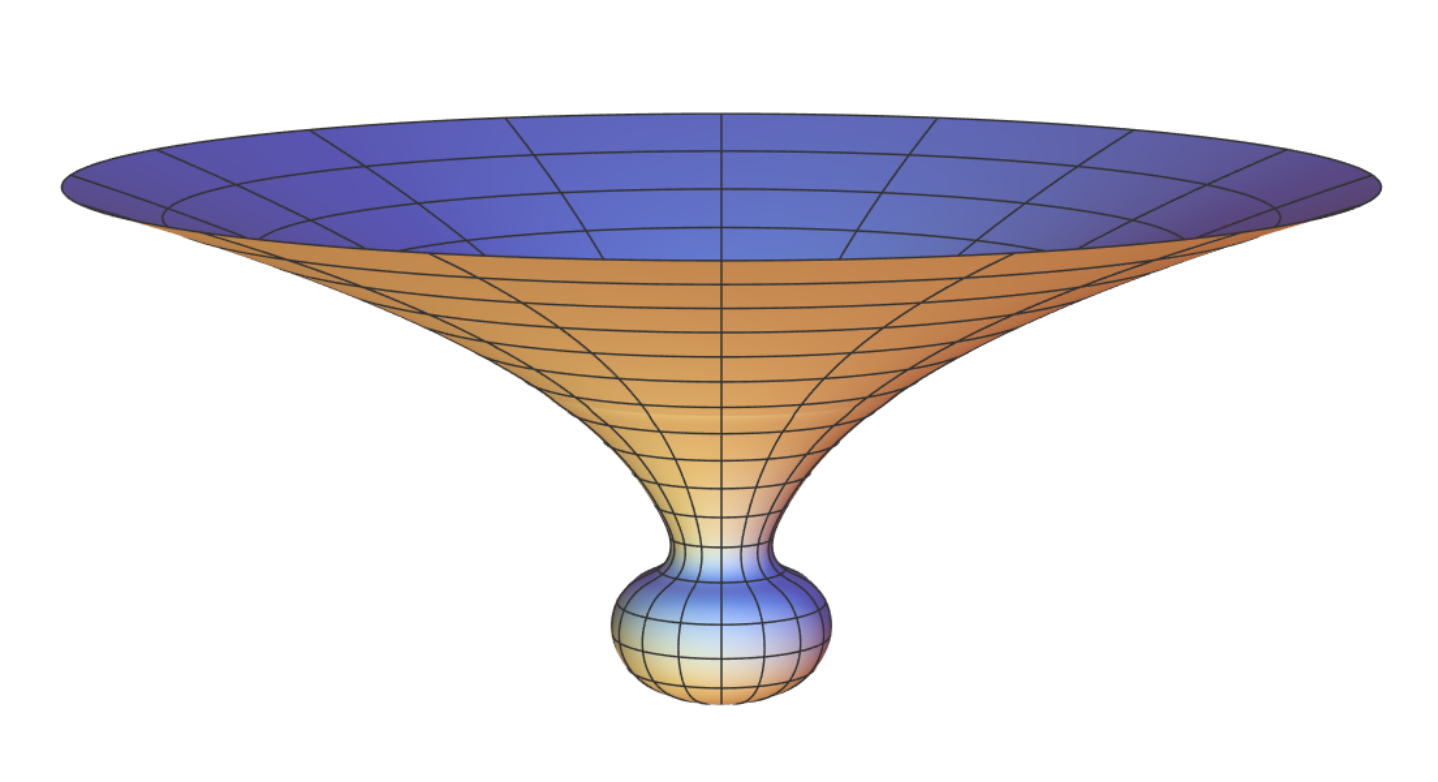}
            \begin{tabular}{ p{0.315\linewidth}  p{0.315\linewidth}
            p{0.315\linewidth}}
\centering (a) & \centering (b) & \centering (c)
\end{tabular}
  		\caption{Embedding diagrams of the late-time spacetime geometry at (a) $t=83.7$, (b) $t=84.6$ and (c) $t=85.5$. }
		\label{fig:plot4}
    \end{figure*}

\begin{figure}[t]
 \centering
		\includegraphics[width=0.8\linewidth]{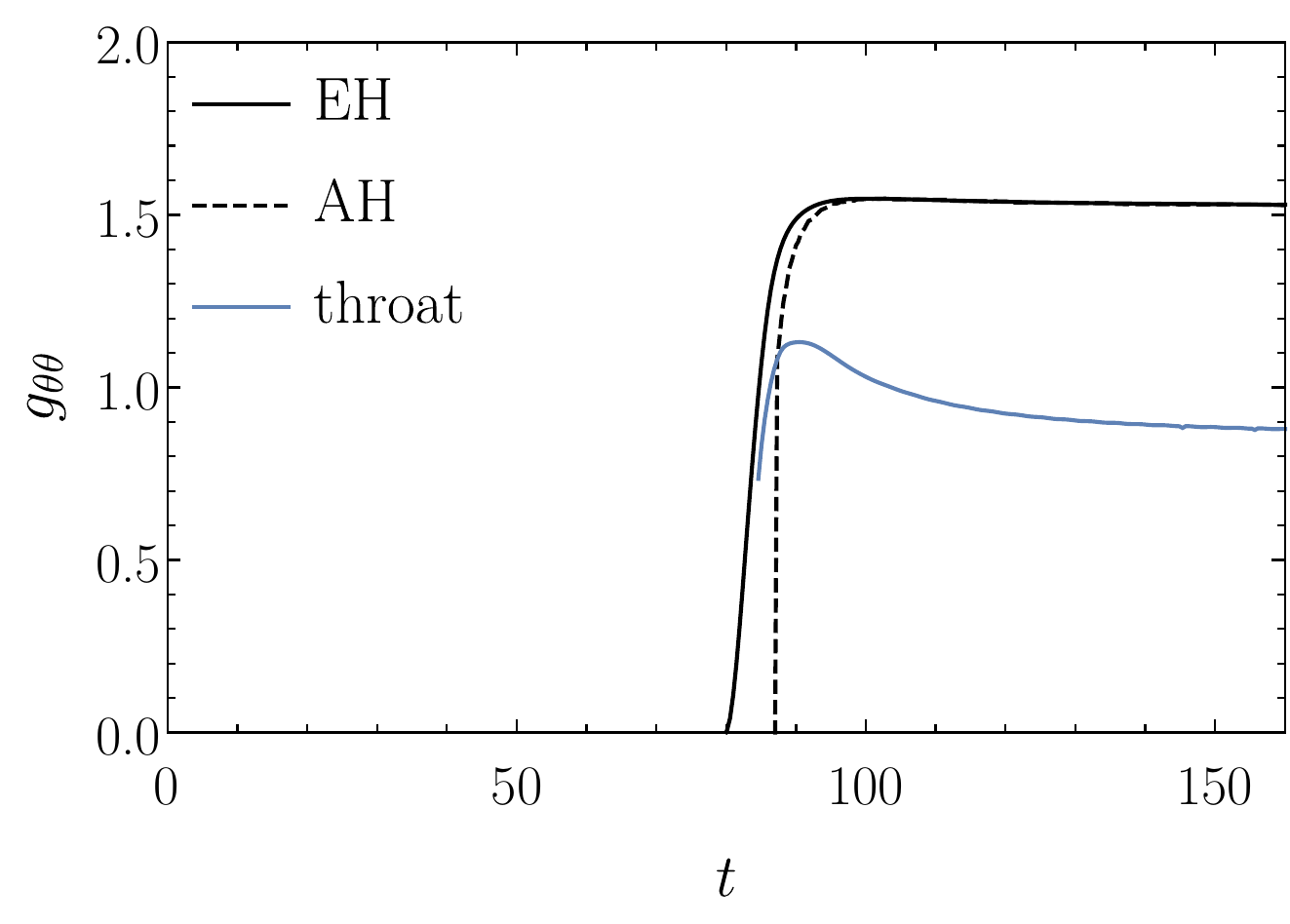}
		\caption{ Time evolution of the location of the throat, AH, and EH in the $f(\R)$ frame using $g_{\theta \theta}$ as pseudocoordinate.}
		\label{fig:plot3}
	\end{figure}

Figure~\ref{fig:plot4} shows embedding diagrams of the late-time spacetime geometry for three illustrative snapshots. They have been computed following the procedure described in~\cite{James:2015ima,Hartle:2003}. The diagrams display an infinite PU  connected to a finite  BU through a throat. The bubble observed at the bottom part of the diagrams corresponds to the BU and its size grows exponentially with time. 
The time evolution of the position of the throat, AH, and EH in the $f(\R)$ frame is displayed in figure~\ref{fig:plot3}. 
The EH appears at $t\approx 80$, the throat at $t\approx84.6$, with a nonzero finite area, and the AH at $t\approx87$. Note that the position of the throat initially grows and then decreases towards an asymptotic value of $g_{\theta \theta} \approx 0.89$. This is a consequence of the slicing employed in the simulation since $g_{\theta \theta}$ is calculated in terms of $q_{\theta \theta}$ and, as mentioned before, the area of the two-sphere does not cover the whole domain. In practice, $g_{\theta \theta}$ approaches the smallest value of $q_{\theta \theta}$ available in the simulation. Since the area of the minimal two-sphere depends directly on the energy density of the scalar field, the slow absorption of the external scalar cloud indicates that it will eventually shrink to zero, closing the umbilical chord connecting the two universes.
The evolution reveals that the throat is always hidden inside the EH, preventing light rays emitted at the BU from escaping to the exterior of the PU. Accordingly, distant external observers will not be able to tell if the outcome of the collapse is an ordinary black hole or a black hole with an inner expanding universe.

\section{Final remarks} Our analysis of the gravitational collapse of boson stars in a metric-affine modified gravity scenario indicates that new dynamics able to trigger dramatic deformations of the space-time structure may be excited at very high energy densities. We have seen that a small patch of space can inflate giving rise to an exponentially growing baby universe. In our model, this occurs in parallel with the development of an apparent horizon, making the internal process analogous to a cosmic bounce and preventing its observation by external observers.  Our results are robust and persist for all values of the gravitational coupling parameter $\xi$ and for other scalar field central amplitudes $\Phi_0$ as long as they are in the unstable branch and the perturbation is high enough to excite the gravitational collapse. 
The fact that metric-affine theories lead to cosmic bounces quite generically suggests that other forms of matter, such as unstable neutron stars, and other gravity theories could lead to outcomes similar to those presented here\footnote{Cosmic bounces may also occur in theories in which the relation between the metrics $g_{\mu\nu}$ and $q_{\mu\nu}$ is not conformal \cite{BeltranJimenez:2017doy,Barragan:2010qb}, which lie beyond the $f(\R)$ family. }. In this sense, we note that the density-dependent modified dynamics of Palatini theories is also present in some instances of scalar-tensor theories of the Horndeski type (compare \cite{Sakstein:2015zoa} and \cite{Olmo:2019qsj}). This suggests that the phenomenology that we find here in the Palatini $f(\R)$ framework could also be present in other relevant gravity theories, which deserves further independent analysis.

 We have seen that the throat area depends on the infalling energy density and shrinks as the external quasi-stationary scalar cloud is absorbed, suggesting that it will eventually close. However, numerical limitations challenge the analysis of this late-time behavior. The case of a stationary spinning black hole with scalar hair in equilibrium, formed through superradiance~\cite{Herdeiro:2014goa,East:2017ovw} or mergers of bosonic stars~\cite{Sanchis-Gual:2020mzb}, could help stabilize the area of the throat, shedding light in this direction.  On the other hand, in the absence of a horizon, a wormhole-like structure could be formed instead. 
 These are aspects to be explored in the future.

Further research on the properties of the BU is necessary to better understand if the inflating phase could be compatible with the mechanism that supposedly contributed to the homogeneity of our own universe in its earliest stages. In this sense, we note that the period in which the scalar energy density builds up at the center of the star seems to provide, within a canonical 4-dimensional picture \cite{Pourhasan:2013mqa}, natural conditions to homogenize the expanding matter. Gravitational collapse in asymmetric bouncing models, such as those emerging from loop quantum gravity \cite{Delhom:2023xxp}, are also worth exploring, as they could alter the post-bounce dynamics potentially leading to {\it black hole to white hole} transitions \cite{Gambini:2020nsf} or black bounce solutions \cite{Gambini:2020nsf,Simpson:2018tsi,Lobo:2020ffi}.  

\acknowledgments

AMF is supported by the Spanish Ministerio de Ciencia e Innovaci\'on with the PhD fellowship PRE2018-083802. NSG is supported by the Spanish Ministerio de Universidades, through a María Zambrano grant (ZA21-031) with reference UP2021-044, funded within the European Union-Next Generation EU.
This work is also supported by the Spanish Agencia Estatal de  Investigaci\'on (grants PID2020-116567GB-C21 and 
PID2021-125485NB-C21 funded by MCIN/AEI/10.13039/501100011033 and ERDF A way of making Europe) and by the project PROMETEO/2020/079 (Generalitat Valenciana). Further support is provided by the EU's Horizon 2020 research and innovation (RISE) programme H2020-MSCA-RISE-2017 (FunFiCO-777740) and  by  the  European Horizon  Europe  staff  exchange  (SE)  programme HORIZON-MSCA-2021-SE-01 (NewFunFiCO-101086251).

\appendix
\section{Note on the numerical framework}\label{appendix}
The BSSN evolution equations are solved numerically using a second-order, partially-implicit Runge-Kutta scheme \cite{Cordero:2012,Cordero:2014}. This scheme can handle in a satisfactory way the singular terms that appear in the evolution equations due to our choice of slicing and coordinates. Explicit details about our numerical implementation have been reported in e.g.~\cite{Sanchis-Gual:2015lje}. In the 3+1 BSSN formalism~\cite{Baumgarte:1998te,Shibata:1995we} space-time is foliated by a family of spatial hypersurfaces $\Sigma_t$ labeled by its time coordinate $t$. We denote the (future-oriented) unit normal timelike vector  of each hypersurface by $n^{\alpha}=(1/\alpha,-\beta^i/\alpha)$, and its dual by $n_{\alpha}=(-\alpha,0,0,0)$, where $\alpha$ is the lapse function and $\beta^i$ is the shift vector. Since the system we study has spherical symmetry, the metric in the Einstein frame reads
\begin{equation}\label{eq:EinsteinMetric}
    \begin{aligned}
        ds_{\text{EF}}^2=&-(\alpha^2 -\beta^x \beta_{x})dt^2+2\beta_{x}dx dt \\
        &+ e^{4 \chi(t,x)}\left( a(t,x)dx^2+x^2 b(t,x) d\Omega^2\right)\,\,,\\
    \end{aligned}
\end{equation}
where $x$ is a radial coordinate, $d\Omega^{2}=d\theta^2+\sin^2\theta d\varphi^2$, $a(t,x)$ and $b(t,x)$ are the conformal metric components, and $\chi(t,x)$ is a conformal factor defined by
\begin{equation}
    \chi=\frac{1}{12}\ln(\gamma/\hat{\gamma}) \,\,.
\end{equation}
Here, $\gamma$ is the determinant of the spacelike metric induced on every hypersuface $\Sigma_t$,
\begin{equation}
    \gamma_{\alpha \beta}=q_{\alpha \beta}+n_{\alpha}n_{\beta} \,\, ,
\end{equation}
and $\hat{\gamma}$ is the determinant of the conformal metric. The latter relates to the full 3-metric by
\begin{equation}
    \hat{\gamma}_{ij}=e^{-4 \chi}\gamma_{ij}\,\,.
\end{equation}

Initially, the determinant of the conformal metric fulfills the condition that it equals the determinant of the flat metric in spherical coordinates, $\hat{\gamma}(t=0)=x^{4}\sin^2\theta$. Moreover, we impose the so-called ``Lagrangian'' condition, $\partial_{t}\hat{\gamma}=0$.

In the BSSN formalism the evolved fields are the conformally related 3-dimensional metric components $a$ and $b$,  the conformal exponent $\chi$, the trace of the extrinsic curvature $K$, the independent component of the traceless part of the conformal extrinsic curvature, $A_a\equiv A^x_{\, x}$, $A_b\equiv A^\theta_{\, \theta}=A^\varphi_{\, \varphi}$, and the radial component of the conformal connection functions $\hat{\Delta}^x\equiv\hat{\gamma}^{mn}(\hat{\Gamma}^x_{mn}-\hat{\Gamma}^x_{mn}(t=0))$ \cite{Escorihuela-Tomas:2017uac,Montero:2012yr}. Explicitly, the BSSN evolution system reads
\begin{equation}
    \begin{aligned}
        \partial_t a= \beta^x\partial_x a + 2a \partial_x\beta^x-\frac{2}{3} a \hat{\nabla}_x \beta^x-2\alpha a A_a \quad,
    \end{aligned}
\end{equation}
\begin{equation}
    \begin{aligned}
        \partial_t b= \beta^x\partial_x b + 2b \frac{\beta^x}{x}-\frac{2}{3} b \hat{\nabla}_x \beta^x-2\alpha b A_b \quad,
    \end{aligned}
\end{equation}
\begin{equation}
    \begin{aligned}
        \partial_t \chi= \beta^x\partial_x \chi +\frac{1}{6} \left(\alpha K-\hat{\nabla}_x \beta^x\right) \quad ,
    \end{aligned}
\end{equation}
\begin{equation}
    \begin{aligned}
        \partial_t K= \beta^x\partial_x K - \nabla^2 \alpha+ \alpha (A_a^2+2A_b^2 +\frac{1}{3}K^2)+4\pi \alpha \left( \rho +S_a+2S_b\right)\quad ,
    \end{aligned}
\end{equation}
\begin{equation}
    \begin{aligned}
        \partial_t A_a= \beta^x\partial_x A_a -\left(\nabla^x\nabla_x\alpha-\frac{1}{3}\nabla^2 \alpha\right)+\alpha \left(R^x_x-\frac{1}{3}R\right)+a K A_a-16 \pi \alpha (S_a-S_b)\quad,
    \end{aligned}
\end{equation}
\begin{equation}
    \begin{aligned}
        \partial_t \hat{\Delta}^x=& \beta^x\partial_x \hat{\Delta}^x -\hat{\Delta}^x\partial_x\beta^x+\frac{1}{a}\partial^2_x\beta^x +\frac{2}{b}\partial_x\left(\frac{\beta^x}{x}\right)+\frac{1}{3}\left(\frac{1}{a}\partial_x(\hat{\nabla}_x \beta^x)+2\hat{\Delta}^x\hat{\nabla}_x\beta^x\right)\\
        &-\frac{2}{a}\left(A_a\partial_x\alpha+\alpha \partial_x A_z\right)+2 \alpha \left(A_a \hat{\Delta}^x-\frac{2}{x b}(A_a-A_b)\right)\\
        &+\frac{2 \alpha}{a}\left[\partial_x A_a-\frac{2}{3}\partial_xK+6 A_a \partial_x \chi\right.\left.+(A_a-A_b)\left(\frac{2}{x}+\frac{\partial_x b}{b}\right)-8 \pi j_x\right] \quad .
    \end{aligned}
\end{equation}

When performing the time evolution of the above functions we have to specify a stress-energy tensor and its 3+1 projections. The case we are concerned with is a boson star in Palatini $f(\mathcal{R})=\mathcal{R}+\xi\mathcal{R}^2$ gravity. Therefore, following \cite{Maso-Ferrando:2021ngp} we write the corresponding stress-energy tensor in the Einstein frame as 
\begin{equation}\label{tensorem}
    \begin{aligned}
        T_{\mu \nu}=&-\frac{2}{\sqrt{-q}}\frac{\partial(\sqrt{-q} K(Z,\Phi))}{\partial q^{\mu \nu}}\\
        =&\frac{1}{2(1+4\xi \kappa \mu^2 |\Phi|^2)}\left[\partial_{\mu}\Phi^*\partial_{\nu}\Phi +\partial_{\nu}\Phi^*\partial_{\mu}\Phi -q_{\mu \nu}\partial^{\alpha}\Phi^*\partial_{\alpha}\Phi -\mu^2 q_{\mu \nu} |\Phi|^2 \right. \\
        &\left. - 2\xi \kappa  \partial^{\alpha}\Phi^*\partial_{\alpha} \left(\partial_{\mu}\Phi^*\partial_{\nu}\Phi +\partial_{\nu}\Phi^*\partial_{\mu}\Phi \right)+\xi \kappa q_{\mu \nu } \partial^{\alpha}\Phi^*\partial_{\alpha}\Phi\partial^{\beta}\Phi^*\partial_{\beta}\Phi\right]\quad.
    \end{aligned}
\end{equation}
 
The projections are performed using the unit normal vector $n^\alpha$ and the induced metric $\gamma^{\alpha \beta}$. The matter source terms appearing in the BBSN evolution equations are:
	\begin{equation}
		\begin{aligned}
			\rho=&n^{\mu}n^{\nu}T_{\mu \nu}\\
			=&\frac{1}{2(1+4\kappa \xi \mu^2 \Phi^2 )}\left[\Pi^2+\frac{\Psi^2}{ae^{4\chi}}+\mu^2\Phi^2 -\kappa \xi\left(\frac{\Psi^2}{ae^{4\chi}}\right)^2+3\kappa\xi\Pi^4-2\kappa \xi \frac{\Psi^2}{ae^{4\chi}}\Pi^2 \right]\quad ,\\ 
		\end{aligned}
	\end{equation}
	
	\begin{equation}
		\begin{aligned}
			S_{a}=&\gamma^{x \mu}T_{x \mu}\\
			=&\frac{1}{2(1+4\kappa \xi \mu^2 \Phi^2 )}\left[\Pi^2+\frac{\Psi^2}{ae^{4\chi}}-\mu^2\Phi^2  -3\kappa \xi\left(\frac{\Psi^2}{ae^{4\chi}}\right)^2+\kappa\xi\Pi^4+2\kappa \xi \frac{\Psi^2}{ae^{4\chi}}\Pi^2 \right]\quad,\\
		\end{aligned}
	\end{equation}
	
	\begin{equation}
		\begin{aligned}
			S_{b}=&\gamma^{\theta \mu}T_{\theta \mu}\\
			=&\frac{1}{2(1+4\kappa \xi \mu^2 \Phi^2 )}\left[\Pi^2-\frac{\Psi^2}{ae^{4\chi}}-\mu^2\Phi^2 +\kappa \xi\left(\frac{\Psi^2}{ae^{4\chi}}\right)^2+\kappa\xi\Pi^4-2\kappa \xi \frac{\Psi^2}{ae^{4\chi}}\Pi^2 \right]\quad,\\
		\end{aligned}
	\end{equation}
	
	\begin{equation}
		\begin{aligned}
			j_{x}=&-\gamma_{x }^{\mu}n^{\nu}T_{\mu \nu}\\
			=&\frac{1}{2(1+4\kappa \xi \mu^2 \Phi^2 )}\left[\frac{1}{a e^{4\chi}}\left(\Pi\Psi^*+\Pi^*\Psi\right)+\frac{2\kappa \xi \Psi^2}{a^2 e^{8\chi}}\left(\Pi\Psi^*+\Pi^*\Psi\right)-\frac{2\kappa \xi \Pi^2}{a e^{4\chi}}\left(\Pi\Psi^*+\Pi^*\Psi\right)\right]\quad .
		\end{aligned}
	\end{equation}

  Correspondingly, the equations of motion for the scalar field are obtained by reformulating the Klein-Gordon equation in terms of the following two first-order variables 
	\begin{eqnarray}
		\Psi& :=& \partial_{x}\Phi\,\, ,\\
			\Pi &:=& n^{\alpha}\partial_{\alpha}\Phi 
			=\frac{1}{\alpha}\left( \partial_{t}\Phi-\beta^{x}\Psi\right)\,\, .
	\end{eqnarray}

 In this way the equations of motion for the scalar field read
	\begin{eqnarray}
		\partial_{t}\Phi&=&\beta^{x}\partial_{x}\Phi+\alpha \Pi\,\,,\\
		\partial_{t}\Psi&=&\beta^{x}\partial_{x}\Psi+\Psi\partial_{x}\beta^x+\partial_{x}\left(\alpha \Pi\right)\,\,,\\
		\partial_{t}\Pi&=&\frac{1-2\kappa \xi Z+\kappa \xi |\Pi|^2}{1-2\kappa \xi Z+2\kappa \xi |\Pi|^2}\left\{\Xi-\frac{\kappa \xi\Pi^2 \Xi^*}{1-2\kappa \xi Z+\kappa \xi |\Pi|^2}\right\}\,\,,
  \nonumber \\
	\end{eqnarray}
where we have introduced the new variable $\Xi$ in order to simplify the notation, defined as 
	\begin{align}
        \Xi:=&\beta^{x}\partial_{x}\Pi+\frac{\Psi}{ae^{4\chi}}\partial_{x}\alpha+\frac{\alpha}{a e^{4\chi}}\left[\partial_{x}\Psi+\Psi\left(\frac{2}{x}-\frac{\partial_{x}a}{2a}+\frac{\partial_{r}b}{b}+2\partial_{x}\chi\right) \right]+\alpha K \Pi \nonumber \\
        &-\frac{\alpha \mu^2 \Phi}{1-2\kappa \xi Z}  +\frac{\alpha \left(Z-\kappa \xi Z^2+\mu^2|\Phi|^2\right)4\xi\kappa\Phi\mu^2}{\left(1+4\kappa \xi \mu^2 |\Phi|^2\right)\left(1-2\kappa \xi Z\right)} \nonumber \\
        &-\frac{4\kappa \xi \mu^2 \alpha}{1+4\kappa \xi \mu^2 |\Phi|^2}\left[-\frac{\Pi}{\alpha}\left(\partial_{t}\Phi^*\Phi+\Phi^*\partial_{t}\Phi\right)+\left(\frac{\Psi}{e^{4\chi}a}+\frac{\Pi\beta^x}{\alpha}\right)\left(\partial_{x}\Phi^*\Phi+\Phi^*\partial_{x}\Phi\right)\right]\nonumber \\
        &-\frac{\alpha \kappa \xi}{1-2\kappa \xi Z}\left[-\frac{\left(\partial_{t}\Psi^*\Psi+\Psi^*\partial_{t}\Psi\right)e^{4\chi}a-|\Psi|^2\left(4e^{4\chi}a\partial_{t}\chi+e^{4\chi}\partial_{t}a\right)}{e^{8\chi}a^2}\frac{\Pi}{\alpha}\right.\nonumber \\
        &\left.-\frac{\left(\partial_{x}\Psi^*\Psi+\Psi^*\partial_{x}\Psi\right)e^{4\chi}a-|\Psi|^2\left(4e^{4\chi}a\partial_{x}\chi+e^{4\chi}\partial_{x}a\right)}{e^{8\chi}a^2}\left(\frac{\Psi}{e^{4\chi}a}+\frac{\Pi\beta^x}{\alpha}\right)\right.\nonumber \\
        &\left.-\left(\frac{\Psi}{e^{4\chi}a}+\frac{\Pi\beta^x}{\alpha}\right)\left(\partial_{x}\Pi^*\Pi+\Pi^*\partial_{x}\Pi\right)\right] \,.
		\end{align} 
 
Within the BSSN formalism we have gauge freedom to choose the ``kinematical variables'', i.e.~the lapse function and the shift vector. As customary in numerical relativity, we choose the so-called ``non-advective 1+log'' condition for the lapse function \cite{Bona:1997hp}, and a variation of the ``Gamma-driver'' condition for the shift vector \cite{Alcubierre:2002kk,Alcubierre:2011pkc},
\begin{equation}\label{eq:gauge}
    \begin{aligned}
        \partial_t \alpha &=-2 \alpha K\quad,\\
        \partial_t B^x&=\frac{3}{4}\hat{\Delta}^x\quad,\\
        \partial_t \beta^x&=B^x \,\,.
    \end{aligned}
\end{equation}
 We also provide the explicit form of the conformal factor $f_\R$. From the Einstein field equations of the Palatini quadratic $f(\R)$ model it can be shown that $\R=-\kappa T$. Therefore,
 	\begin{equation}
			f_{\R}=1+2\xi\kappa\R
			=\frac{1-8 \kappa \xi V}{1-2\kappa \xi Z}\,\,.
	\end{equation}

\begin{figure}[t!]
    \centering
    \includegraphics[width=0.75\linewidth]{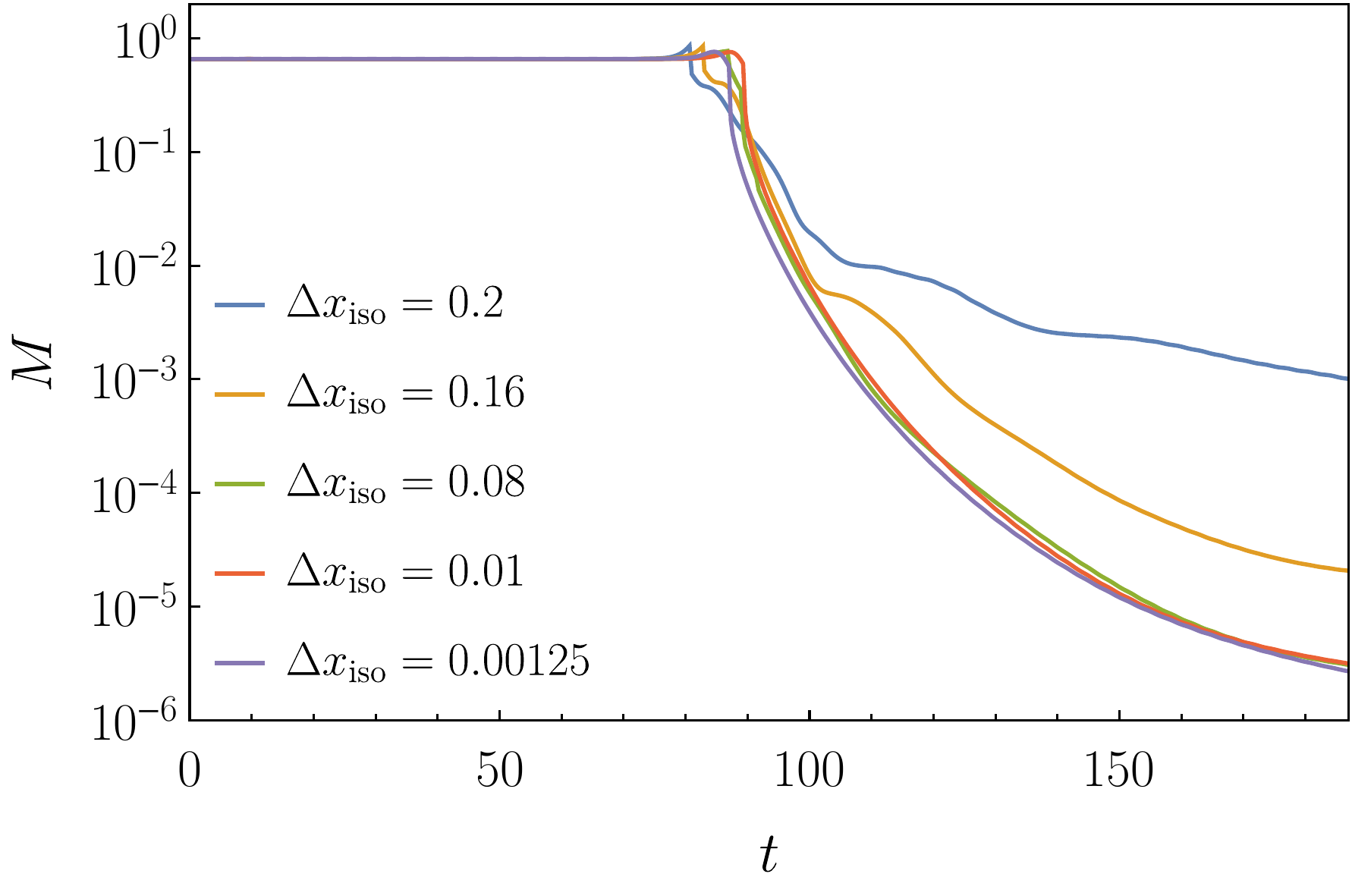}
    \caption{Time evolution of the total mass for different isotropic grid resolutions with fixed polar-areal grid resolution $\Delta x_{\text{pa}}=0.0025$. The finest isotropic grid (black curve) is the one used in the simulations discussed in the main text.}
    \label{fig:mKG}
\end{figure}

	In addition to the evolution equations, the Einstein-Klein-Gordon system also contains the Hamiltonian and momentum constraint equations. 
 These equations read
 \begin{equation}
		\mathcal{H}\equiv R-(A_{a}^2+2A_{b}^2)+\frac{2}{3}K^2-2\kappa\rho=0\quad,
	\end{equation}
	\begin{equation}
		\begin{aligned}
			\mathcal{M}_{x}\equiv& \partial_{x}A_{a}-\frac{2}{3}\partial_{x}K+6A_{a}\partial_{x}\chi+(A_{a}-A_{b})\left(\frac{2}{x}+\frac{\partial_{x}b}{b}\right)-\kappa j_{x}=0 \quad.
		\end{aligned}
	\end{equation}

\begin{figure}[h]
    \centering
    \includegraphics[width=0.75\linewidth]{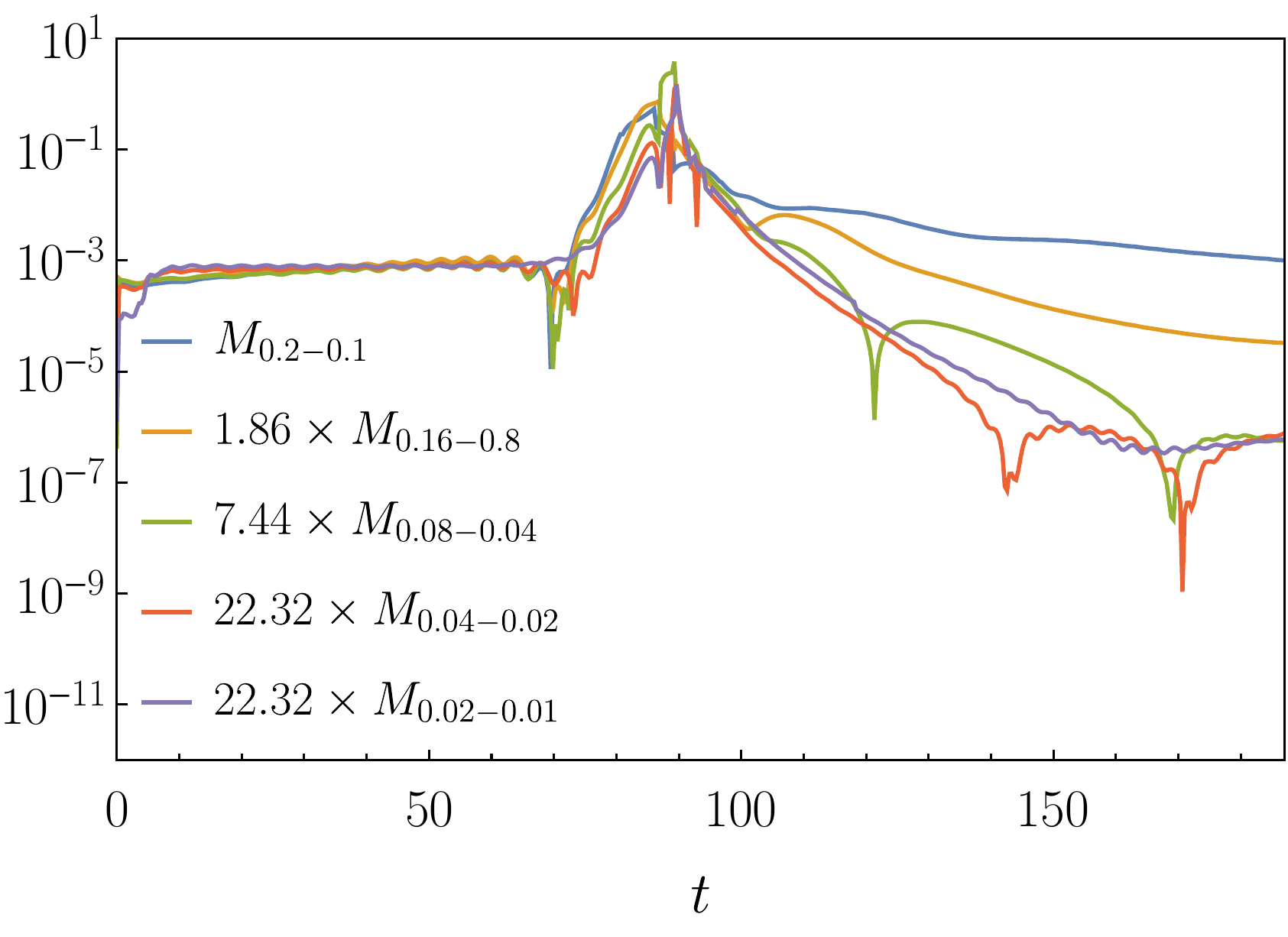}
    \includegraphics[width=0.75\linewidth]{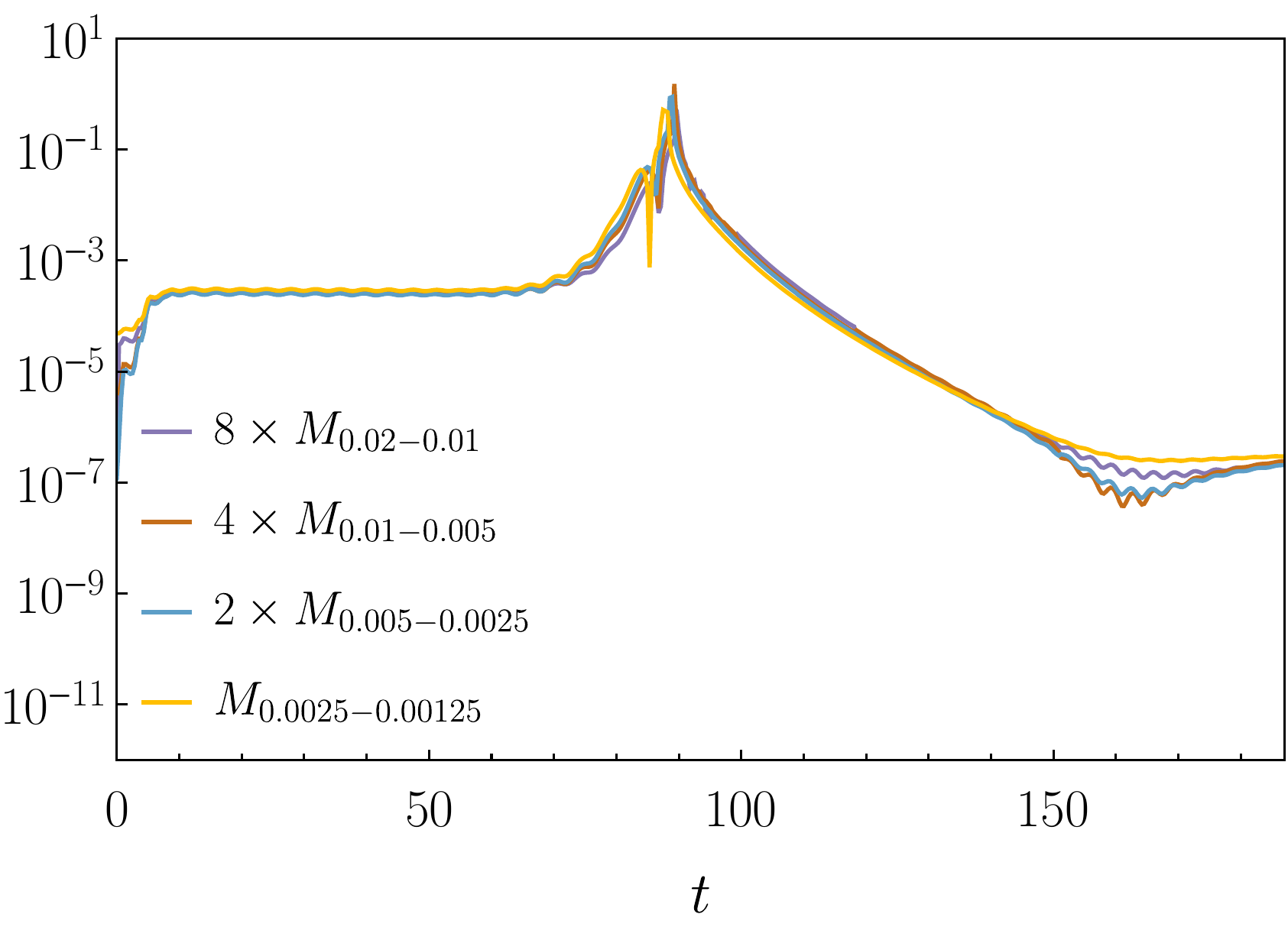}
    \caption{Comparison of the time evolution of the difference between total masses calculated for different isotropic grid resolutions with fixed polar-areal grid resolution $\Delta x_{\text{pa}}=0.0025$. }
    \label{fig:convergence}
\end{figure}

\begin{figure}[h]
    \centering \includegraphics[width=.75\linewidth]{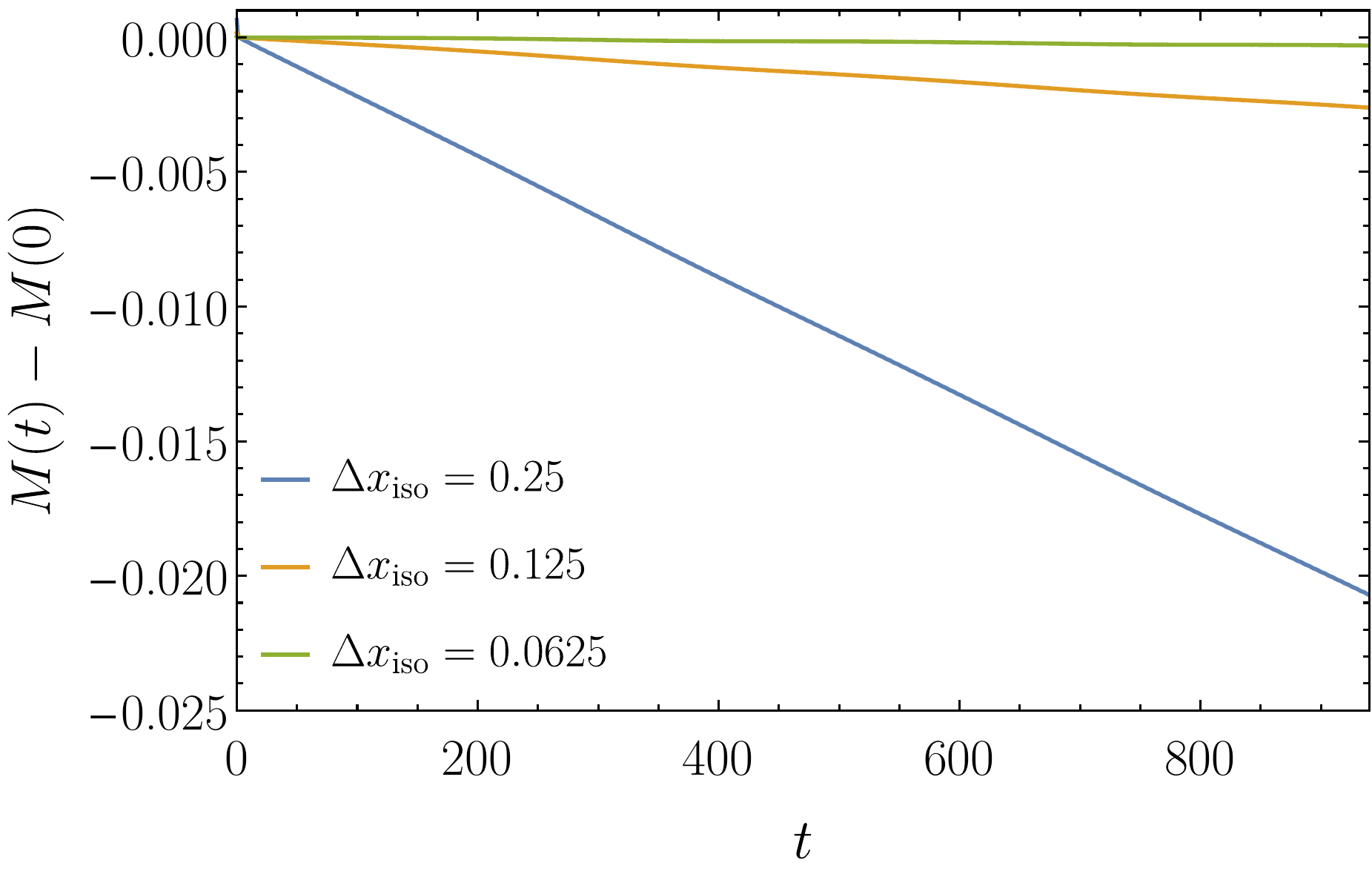}
    \includegraphics[width=.75\linewidth]{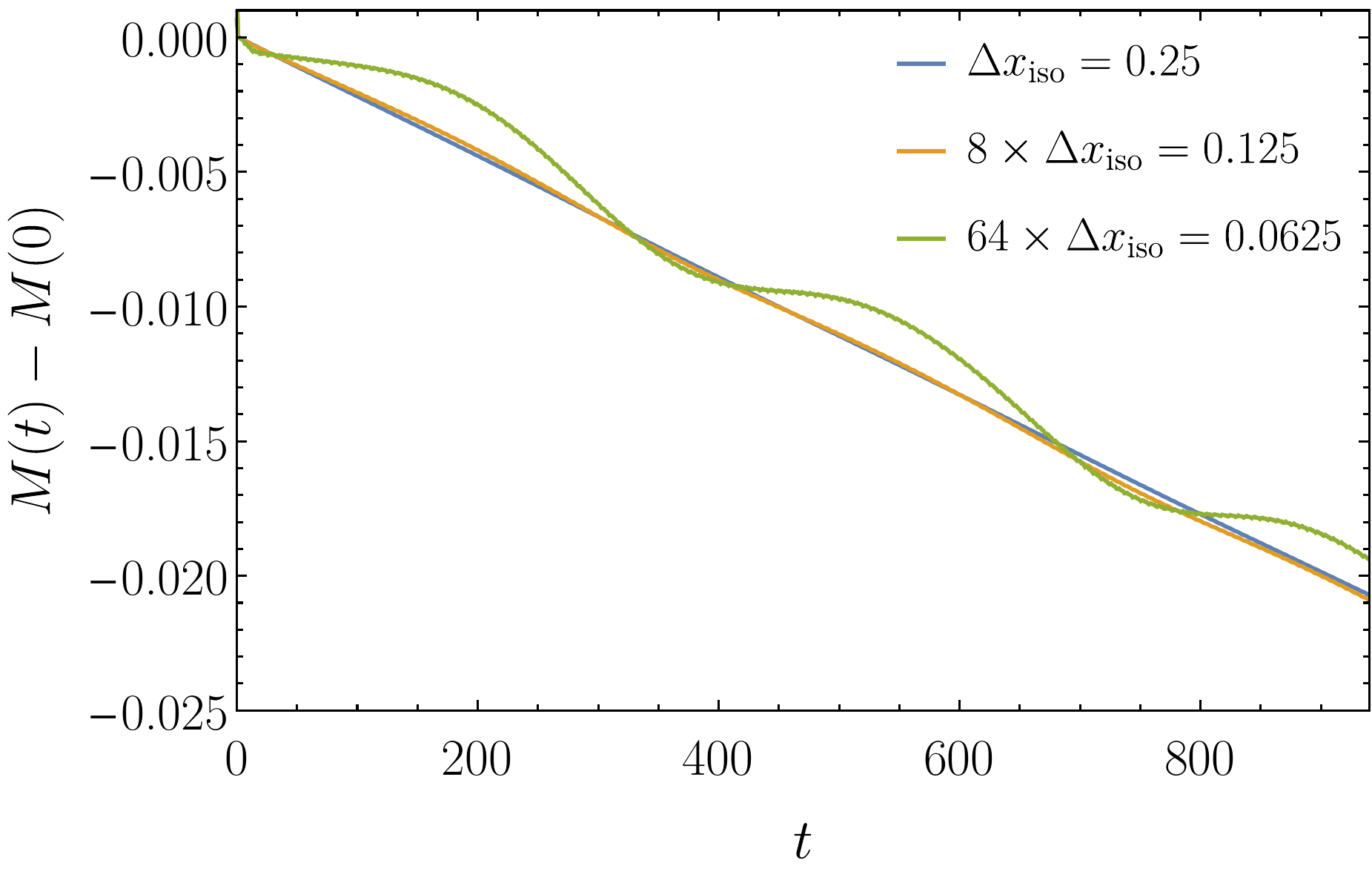}
    \caption{Time evolution of the total mass of a stable boson star model. Top panel: Difference of the instantaneous total mass and its initial value for three different evolution grid resolutions $\Delta x_{\text{iso}}$. Bottom panel: The quantities of the top panel are rescaled to show third-order convergence.}
    \label{fig:stable}
\end{figure}

The total mass of the spacetime can be calculated by integrating the stress-energy tensor at each spatial hypersurface $\Sigma$ \cite{Herdeiro:2016tmi}
\begin{equation}\label{eq:massa_kg}
  M = \int_{\Sigma} 
\left(2T_t^t-T_\mu^\mu\right)\alpha\sqrt{\gamma} \,dr\,d\theta\, 
d\varphi\ .
\end{equation}
Figure~\ref{fig:mKG} shows the time evolution of the total mass for several isotropic grid resolutions. In order to check the convergence of the results, masses calculated from different grid resolutions are compared according to 
\begin{equation}
    M_{u-v} = |M(\Delta x_{\text{iso}}=u)-M(\Delta x_{\text{iso}}=v)| \quad.
\end{equation}
Setting $\Delta x_{\text{pa}}=0.0025$ (the spatial resolution needed in the polar-areal grid used to compute the initial data) and choosing several resolutions for the isotropic grid, namely from $\Delta x_{\text{iso}}=0.2$ to $\Delta x_{\text{iso}}=0.04$, we find second-order convergence during the early contraction phase and third-order convergence during the collapse and black hole formation phase. This can be inferred by the multiplicative factors employed in the first three curves in the legend of the top panel of Figure~\ref{fig:convergence}. However, increasing the resolution of the isotropic grid from $\Delta x_{\text{iso}}=0.04$ to $\Delta x_{\text{iso}}=0.02$, the convergence order drops to $\sim1.8$ in the early phase. In addition, for even higher resolutions of the isotropic grid, the accuracy of the evolution does not improve (see bottom panel of Figure~\ref{fig:convergence}). In this analysis we are only considering the numerical error coming from the finite-differencing of the differential equations. This dominates the error if we use resolutions coarser than that used to compute the initial data. However, we note that the change of coordinates from polar-areal to isotropic (see details on the specific transformation in~\cite{Escorihuela-Tomas:2017uac}) also introduces an additional source of error, that is reflected in the loss of convergence shown in the bottom panel of  Fig.~\ref{fig:convergence}. In addition, since we do not further change $\Delta x_{\text{pa}}$ in this analysis, increasing the isotropic grid resolution does not lead to an improved convergence for the higher resolution cases discussed here. Despite the lack of convergence for an isotropic grid with $\Delta x_{\text{iso}}=0.00125$, our simulations needed to use such high resolution in order to populate the vicinity of the origin with a sufficiently large number of cells (even though the accuracy of the result does not increase at the expected rate). A remedy to this shortcoming, which we believe does not affect the validity of the findings reported in this work, would be to compute the initial data directly in isotropic coordinates and thus avoid the coordinate transformation for the evolution. Further developments in this direction will be reported elsewhere.  We also note that, similarly, first-order convergence is found for the polar-areal grid when the isotropic grid is fixed.

For completeness, we discuss the code convergence properties when evolving a stable boson star model using the same $f(\mathcal{R})$ theory (with $\xi=0.1$). We select a model with $\omega/\mu=0.9545$, $\Phi(r=0)=0.02$, and total mass $M=0.471$, and a polar-areal grid resolution for the initial data of  $\Delta x_{\text{pa}}=0.00125$. The results are plotted in Figure~\ref{fig:stable}. For this stable model, numerical errors from finite-differencing dominate the evolution and the total mass decreases with a drift that depends on resolution (see top panel of Fig.~\ref{fig:stable}). The rate of convergence of the total mass for this stable model is third order, as shown in the bottom panel of Fig.~\ref{fig:stable}.

\end{document}